\documentclass[conference]{sig-alternate}

\usepackage{color}
\usepackage{graphicx}
\usepackage{textcomp}
\usepackage{array}
\usepackage{xspace}
\usepackage{caption}
\usepackage{subcaption}

\newcommand{\cloudrone}{$\mathit{Cloudrone}$\xspace}

\newif\ifeurasia
\eurasiafalse
\ifeurasia
\fi
\makeatletter
\def\@copyrightspace{\relax}
\makeatother
\begin{document}
\bibliographystyle{plain}
\title{Cloudrone: Micro Clouds in the Sky}

\numberofauthors{6} 
%

%
%
\author{Arjuna Sathiaseelan$^{1}$, Adisorn Lertsinsrubtavee$^{1}$, Adarsh Jagan S$^{2}$, Prakash Baskaran$^{2}$, 
\\Jon Crowcroft$^{1}$ \\
$^{1}$University of Cambridge, \\$^{2}$National Institute of Technology, Trichy \\
as2330@cam.ac.uk, al773@cam.ac.uk, jac22@cam.ac.uk }

\maketitle
\sloppy
\begin{abstract} 
Recent years have witnessed several initiatives on enabling Internet access to the next three billion people. Access to the Internet necessarily translates to access to its services.  This means that the goal of providing Internet access requires access to its critical service infrastructure, which are currently hosted in the cloud. However, recent works have pointed out that the current cloud centric nature of the Internet is a fundamental barrier for Internet access in rural/remote areas as well as in developing regions. It is important to explore (low cost) solutions such as micro cloud infrastructures that can provide services at the edge of the network (potentially on demand), right near the users.

In this paper, we present \cloudrone - a preliminary idea of deploying a lightweight micro cloud infrastructure in the sky using indigenously built low cost drones, single board computers and lightweight Operating System virtualization technologies. Our paper lays out the preliminary ideas on such a system that can be instantaneously deployed on demand. We describe an initial design of the \cloudrone and provide a preliminary evaluation of the proposed system mainly focussed on the scalability issues of supporting multiple services and users.
\end{abstract}

\vspace{-2ex}
\section{Introduction}

Recent years have witnessed several initiatives on enabling Internet access to the next three billion people.  It is widely recognized by all the stakeholders involved in the game of universal service provisioning, that providing access to the Internet has no one size fits all solution for enabling wider universal Internet access but requires exploring a variety of solutions.  This is evident from organizations like Facebook and Google who are on a (noble) mission to connect the next three billion through novel ways utilizing high altitude platforms such as drones, balloons and satellites.

Access to the Internet necessarily translates to access to its services.  This means that the goal of providing Internet access requires access to its critical service infrastructure, which are currently hosted in the cloud. However, recent works have pointed out that the current cloud centric nature of the Internet is a fundamental barrier for Internet access in rural/remote areas as well as in developing regions. 

Taking Africa as an example, we recently conducted a study mapping the African web ecosystem \cite{fanou}. Our study highlighted that majority of the African web infrastructure is placed outside the continent. This has direct impact on aspects of DNS resolution times as well as HTTP page load times causing users to experience higher latencies for downloading the webpages and poor quality of experience. Our paper concluded that there is a need for localised service infrastructures.

Access to  services are also extremely important (even potentially life saving) during humanitarian crisis where access to services are challenged - due to intermittent connectivity, heavy interference and congestion leading to increased latencies to global servers or in most cases no access to them. Hence its not rocket science to understand that access to localized service infrastructures is of paramount importance to solve the global access problem. 

Recent work on mobile edge computing \cite{5280678} aims to push computation right at the edge of mobile networks, enabling computations at the edge improving latencies and performance. The recent work on infrastructure mobility \cite{Gowda:2014:IMW:2670518.2673862} illustrates the interesting concept of making the access infrastructure mobile thus providing better and much more efficient coverage based on the need/demand from the users. So an obvious question or follow up to this exciting idea is rather than making the access infrastructure mobile, why cannot build on such work to make the service infrastructure mobile i.e. rather than expecting the users to access the cloud services, why cannot we directly provision the cloud service infrastructure to the user on demand even in the absence of a terrestrial infrastructure e.g. rural/remote areas or disaster zones?

In this paper, we present \cloudrone - a preliminary idea of deploying a lightweight micro cloud infrastructure in the sky using indigenously built low cost drones, single board computers and lightweight Operating System virtualization technologies such as unikernels/dockers \cite{Madhavapeddy:2013:ULO:2499368.2451167}. Our paper lays out the preliminary ideas on such a system that can be instantaneously deployed on demand.  
\vspace{-2ex}
\section{Cloudrone design}
To enable a lightweight micro cloud infrastructure in the sky using drones, we bring together a couple of fascinating recent innovations in computing: single board computers such as the Raspberry PIs (PI) and lightweightOS virtualisation technologies such as Dockers\footnote{www.docker.com}
and integrate them with lightweight quadcopters. There were two possibilities for us to build the cloudrone: integrate the PI with an off the shelf ready made quadcopter or design and build an entire quadcopter from scratch with the PI powering both the drone as well as acting as a micro cloud server. We decide to go for the latter since we wanted to have the drone as lightweight and low cost as possible. 

\subsection{Building a low cost drone}

\subsubsection{Designing the drone}
There are two main challenges in building a low cost quadcopter. It has to be light weight yet robust and it has to be economical. The cost cannot be controlled on components such as the motors, Electronic Speed Controllers (ESCs) and the battery. So the focus was on the other two major areas to reduce the cost. The frame and the RC controller. For building the frame, some materials like carbon fiber are light and strong but they are expensive compared to other materials like aluminium shafts or high quality reinforced plastic. 
The frame of the quadcopter consists of carbon infused plastic, the landing gear is made of light-weight square aluminium sections fitted with high quality foam sponge balls to withstand crashes. Four brushless DC motors (1400KV, 800gms thrust) connected with carbon fibre propellers (8x4.5") drive the drone.

The speed of the motors is controlled by four Electronic Speed Controllers (Peak Current 30A) with in-built battery eliminator circuits that provide 5V, 2A for powering up other circuits like the KK2 flight controller board and the PI. These ESCs are instructed by control signals that are generated by the output pins of the KK2 flight controller board. 
The KK2 board has a variety of modes to control different UAVs like tricopters, quadcopters, hexacopters and octocopters easily. So any extension to the current quadcopter design can be easily accommodated and realised quickly. The KK2 has input pins which are generally connected to RC receivers. But RC transmitters and receivers tend to be quite expensive and make up a large portion of the entire cost.

A good quality 4 channel RC transmitter costs more than \pounds200. To build a low cost quadcopter that can act as a micro cloud server, we eliminated the RC transmitter and receiver entirely and replaced it with the PI. The PI has two important functions i.e., controlling the drone's movements and to run micro services. Since controlling the quadcopter is now being done by the PI, it can be programmed to fly autonomously along with manual override. 
We used the PI 2 model B equipped with quad-core ARMv7 CPU 900 MHz. The RAM memory is 1024 MB. It requires power feed about 800 mA.  The PI runs an ARM version of Ubuntu 14.04. We used a Edimax WiFi dongle (EW-7811Un) for providing the WiFi interface.


\begin{figure*}[ht] 
  \begin{subfigure}[b]{0.33\linewidth}
    	\centering
    	\includegraphics[width=5.5cm,height=4.2cm]{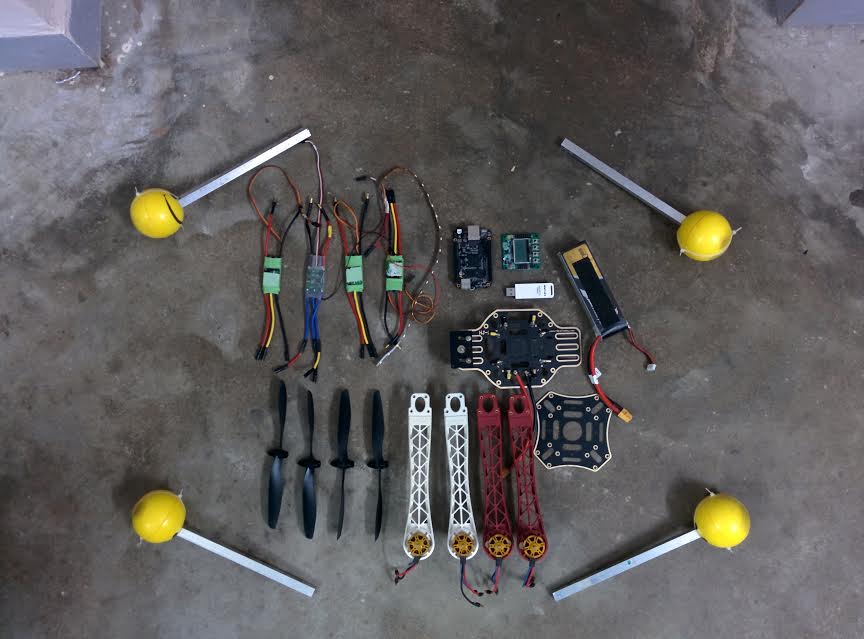} 
    	\caption{\scriptsize{Quadcopter components}} 
    	\label{fig:test1} 
  \end{subfigure}
  \begin{subfigure}[b]{0.33\linewidth}
    	\centering
    	\includegraphics[width=5.5cm,height=4.2cm]{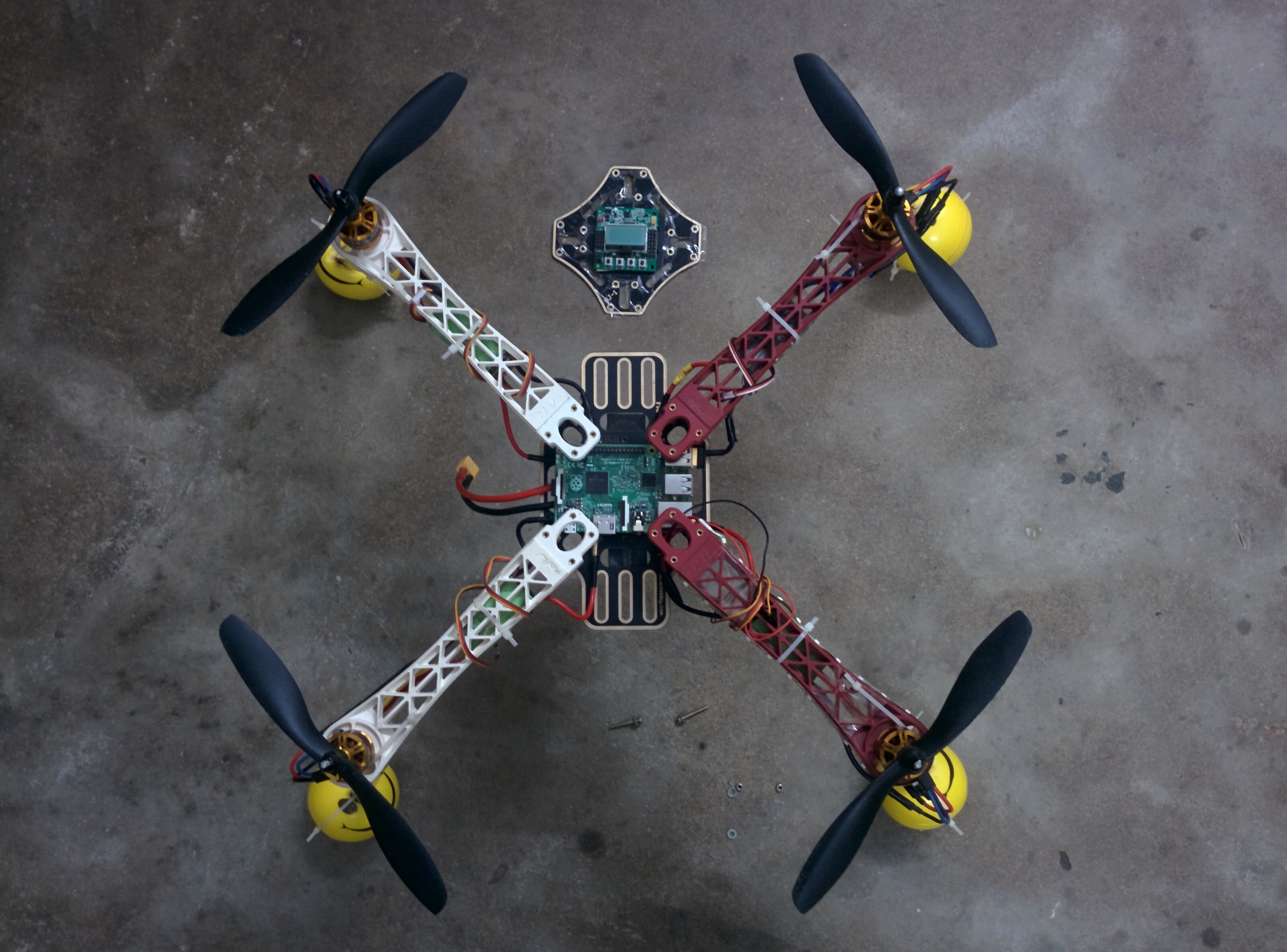} 
    	\caption{\scriptsize{Quadcopter fully assembled}} 
    	\label{fig:test2} 
  \end{subfigure} 
  \begin{subfigure}[b]{0.33\linewidth}
      	\centering
    	\includegraphics[width=5.5cm,height=4.2cm]{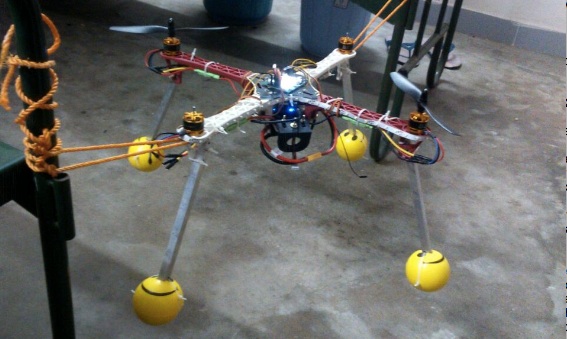}
	 \caption{\scriptsize{Tuning the quadcopter using cot beds}} 
    	\label{quad_tuning}
 \end{subfigure}
  
\vspace{-2ex}
 \caption{Cloudrone Prototype}
\label{fig:Drone}
\end{figure*}

\subsubsection{Controlling the drone}

The PI generates a portable Wi-Fi Access Point to which the laptop is connected. A gaming joystick is connected to the laptop to control the drone manually. By using a python module called Pygame~\footnote{http://www.pygame.org/}, the joystick emulates an RC transmitter with different buttons and 2-axis joysticks, acting as controls for roll, pitch, yaw and throttle. A graphical user interface was also created to monitor the orientation, altitude, battery level, GPS co-ordinates.

The PI and the base laptop are connected by a software framework called ROS (Robot Operating System)~\footnote{http://www.ros.org/} which simplifies sending and receiving data with the concepts of publishing and subscribing data. Data which may be originating from a controller or from a sensor is published onto a ``topic", from which any other component of the entire system can access it by subscribing to that topic. 

In our  design, the joystick values are published to a topic which is subscribed by the PI. The PI then uses that data to send Pulse Width Modulated control signals to the KK2 board. The program that runs on the PI has been designed to give signals exactly as an RC receiver would. So, for the KK2 board the signals appear to be coming from an RC receiver. 

Based on the signal it receives, the KK2 board sends corresponding signals to the ESCs, which in turn send the Brushless DC (BLDC) motors the appropriate currents to change their speeds. All the electronics, motors and ESCs are powered using a Lithium Polymer battery (4200mAh 3S 35C).

In order to maintain stability the KK2 implements P-I (Proportional-Integral) controller. The user has to manually set values to constants that govern the flight control. In the KK2 board, there are two important constants: P-gain, and I-gain for roll, pitch and yaw. These constant values will be the same for both roll and pitch owing to symmetry.

%
Firstly considering only one axis, say roll, setting I-gain to zero, P-gain is increased until the quadcopter produces oscillations periodically. If P-gain is too low, the reaction to roll left or right will be sluggish. If the P-gain is too high the quadcopter produces high frequency oscillations. Based on observations, the P-gain is set to an appropriate value. I-gain is then increased to ensure that the quadcopter does not drift from the set-point. If I-gain is too high, there will be a lot of overshoot from the set point and if it is too low there will be a lot of drift. The quadcopter will not reach the set point forever. In order to tune the roll axis, the quadcopter is tied along the pitch axis to two parallel cots which act as a stable test rig. 

The accelerometers and gyroscopes in the KK2 board are then calibrated. We place the quadcopter on a tested perfectly flat surface and run this calibration on the KK2 board to make sure that the pitch and roll angles are zero while hovering. The ESCs are then calibrated to start at the same time. This is done by pressing the Back and Enter buttons simultaneously while switching on the KK2 board with all the ESCs connected. The joystick connected to the laptop is set to 100\% throttle. The ESCs beep to indicate that they have been calibrated.

Propellers, even the high quality ones may have some weight discrepancies which affect the quadcopter's flight when the motors spin at 14,000 RPM. So with a simple balancing test, we find out which side of the propeller is lighter than the other. To make the weights of both sides of the propeller equal, we stick cellophane tape on the lighter side. 

Similarly to reduce the vibrations generated by the BLDC motors we first measure the vibrations using an android app and stick cellophane tapes accordingly to all four motors and minimize these vibrations as they can corrupt the values of the accelerometer and gyroscope in the KK2 board.

\subsubsection{Benchmarking the drone}
\label{Benchmarking the drone}
The quadcopter's final weight is approximately 1.5 kgs.  The maximum altitude to which it was flown was 50 feet. With the current setup it can fly up to 100 feet theoretically. The effectiveness of the landing gear was also tested. The quad copter was dropped from a 2-storey building (45 feet high free fall) and no components were damaged. 

The quadcopter's flight time has been measured in two different case scenarios. First with a 2200mAh battery with 60C discharge rating and then with a 4200mAh 35C battery. The flight duration and the throttle levels vary noticeably in these two scenarios. 
When we use the 2200mAh battery the payload of the quadcopter was bout 50 gms higher than when the 4200mAh battery was used. These are the observations:
\vspace{-2ex}
\begin{itemize}
\item The quadcopter in the first scenario lifts only at 70\% throttle level and the flight time is very poor owing to the smaller capacity of the battery and slightly heavier load. The estimated and realised flight times are only around 2-3 minutes.
\vspace{-2ex}
\item On redesigning certain aspects of the quadcopter to compensate for the heavier 4200mAh battery, the overall weight of the quadcopter was reduced by 50 gms. The quadcopter now lifts at 55\% throttle and was able to have sustained flight for over 8 minutes. If a smaller battery is used in the lighter quadcopter, the performance is still not as good as using a bigger and higher capacity battery.
\item By using a higher mAh rated battery, the flight time can be further increased to an estimated 15 minutes.
\end{itemize}
\vspace{-1ex}

The quadcopter's CPU consumption was measured  during the flight operation by using \emph{sysstat} tools. On average our quadcopter consumes 30\% for the flight control of the quadcopter leaving 70\% for other processes.

To measure the WiFi coverage of the quadcopter while it was in the air, we measured the signal strength from various distances (with clear line of sight) using Wigle\footnote{https://wigle.net}. We observed signal strengths varying from -65 dBm to -78 dBm within the distances 10 to 50 meters north, east and west of the direction of the WiFi dongle. South being the direction away from the dongle had relatively poor coverage (between -75 dBm to -89 dBm). WiFi coverage can be enhanced by using additional portable lightweight WiFi hotspots that can provide wider coverage (such as the TP-Link MR3040).

\subsection{Integrating the micro cloud with the drone}
Our vision to build a micro cloud infrastructure in the air is to use a swarm of PIs on drones acting as micro cloud servers running Dockers enabling us to run several lightweight containers. The lightweight nature of docker containers significantly reduce the size of image compared to the heavyweight VMs \cite{7214101}. For instance, we can build a minimal static web server with only 2MB size. This can be further reduced with new lightweight OS virtualisation technologies such as IncludeOS
\footnote{http://www.includeos.org}.

To interconnect the PIs as a swarm/mesh, each of these devices also act as mobile routers operating in two WiFi communication modes: one operates in the WiFi ad-hoc mode to allow Optimised Link State Routing (OLSR) protocol, constructing a Mobile Ad hoc Network (MANET)~\cite{manet}; the other operates in the Access Point (AP) mode to allow user devices to connect with DHCP.  The benefit of using OLSR protocol compared to other MANET routing protocols is it uses a special mechanism called Multi-Point Relay (MPR) to reduce the number of flooded messages. Only a few devices that are located in strategically better spatial positions are chosen to relay (i.e., re-transmit) messages in the path from a source device to a destination device. The MPR mechanism helps reduce overall energy consumption. 

To create a swarm of micro cloud servers, we use the Docker swarm technology that allows us to create a cluster of multiple docker hosts (PI running Docker) and migrate the service containers across the cluster. Specifically, there are two types of nodes classified in the swarm. 1. Swarm manager - who manages overall resources (e.g., swarm members, number of running containers) and decides where to place the service containers. 2. Swarm agent - nodes registered with the swarm manager. When the swarm manager and agents are created, they have to register with the discovery backend as members of the swarm. The discovery backend maintains an up-to-date list of swarm members and updates that list with the swarm manager. The swarm manager uses this list to assign tasks and schedule the service containers to the agents. To determine where to place the new container in the swarm, the swarm manager uses either \emph{spread} or \emph{bin packing} strategy to compute the rank regarding node's available CPU, RAM and the number of running containers. With the spread strategy, the swarm manager gives a priority to the node who has the largest available memory or has the minimum number of running containers. On the other hand, the bin packing strategy tries to pack as many containers to a node until reaching its maximum capacity (e.g., RAM, CPU). In this sense, a swarm can optimise the number of nodes running the containers and leave room for future assignment which may require large space of resources.   

\subsection{Deploying the Cloudrone}
\begin{figure}[ht] 
      	\centering
    	\includegraphics[width=0.4\textwidth]{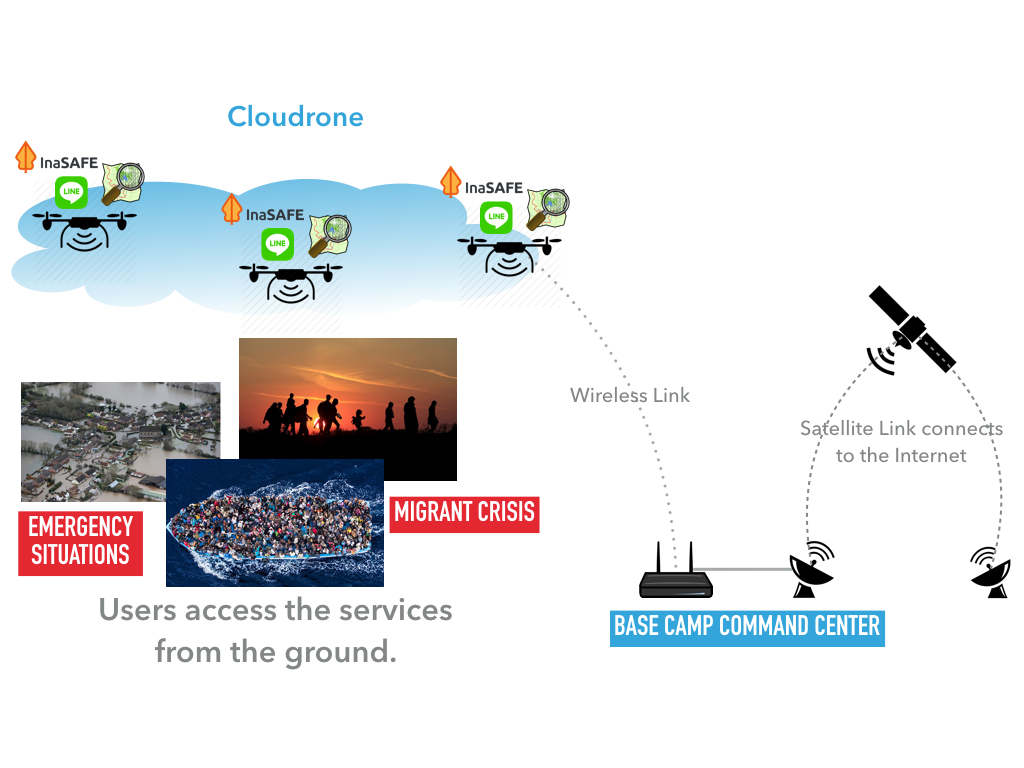}
	\vspace{-2ex}
    	 \caption{An example of Cloudrone's operation} 
    	\label{img:cloudrone}
\end{figure}

\cloudrone targets to provide localised service infrastructure in challenged network environments. Such scenarios refer to the emergency and post-disaster situations wherein traditional communication services are completely inoperable. An example of \cloudrone's deployment is illustrated in Figure~\ref{img:cloudrone}. 

A base camp command center with backhaul Internet connectivity (e.g., satellite link) is set up close to the affected area. A cluster of drones can fly to cover the target area (and could also land), forming a mesh network and provide localised services to the users (immigrants or post-disaster victims) on the ground via WiFi. A variety of (crucial) services (lightweight docker containers) can be either pre-loaded onto the PI or on demand from the ground. The MANET of drones, facilitates the swarm manager to communicate and control the cluster remotely from the base camp through the long haul link. The swarm manager can update the necessary services from the Internet and disseminate throughout the cluster.   
\vspace{-2ex}
\vspace{-2ex}

In some operations, the cluster can be out of contact with the swarm manager (i.e., the target area is far away from the command center). To deal with any interruption of service, we can create a primary swarm manager operating as the main point of contact and multiple replicas to be the backup swarm managers. Using this feature, the replicas can seamlessly take over the functionalities from the primary swarm manager when it fails. If the cluster fails to contact the primary swarm manager, the most powerful replica automatically takes over the control.  

Battery life time is one of the major challenges in our deployment, since the flight time is limited to 15 minutes. To mitigate this problem, we envision that the drone can either fly or land to provide service access. In case, the battery is low, our drone can decide to land on the ground to save the battery while the docker containers and WiFi access point are still operating to provide service access. 

\section{Preliminary Benchmarking}
The preliminary benchmarking presented in this paper focusses on understanding the scalability issues of lightweight OS virtualisation technology such as Docker on a PI. We focus on two main questions 1) how many Docker containers can a single PI support? 2) how many user requests can a single container running on a PI support? 

\begin{figure*}[ht] 
  \begin{subfigure}[b]{0.24\linewidth}
    	\centering
    	\includegraphics[width=1\textwidth]{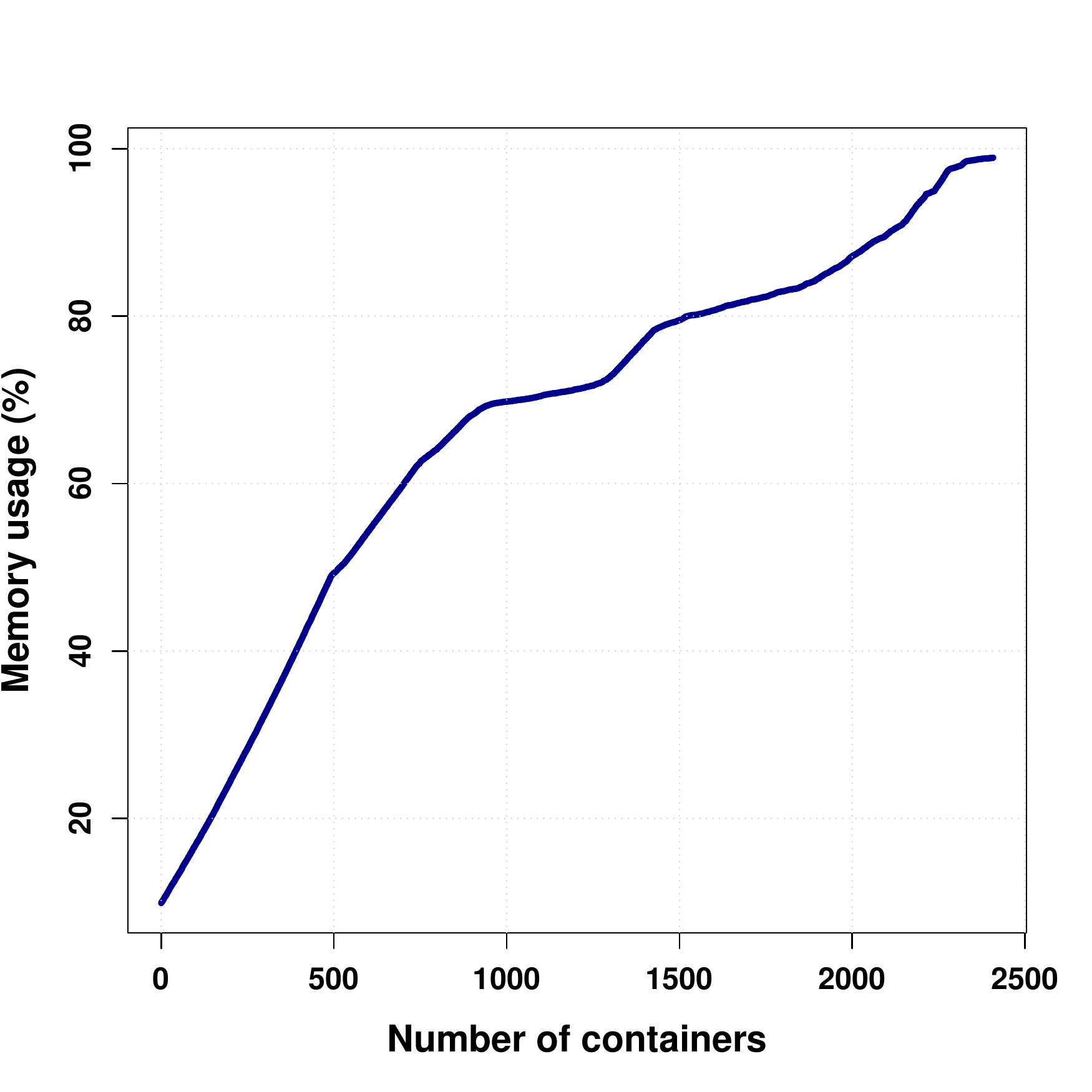}
    	 \caption{\scriptsize{Memory usage}} 
    	\label{result:mem}
	  \end{subfigure}
  \begin{subfigure}[b]{0.24\linewidth}
    	\centering
    	\includegraphics[width=1\textwidth]{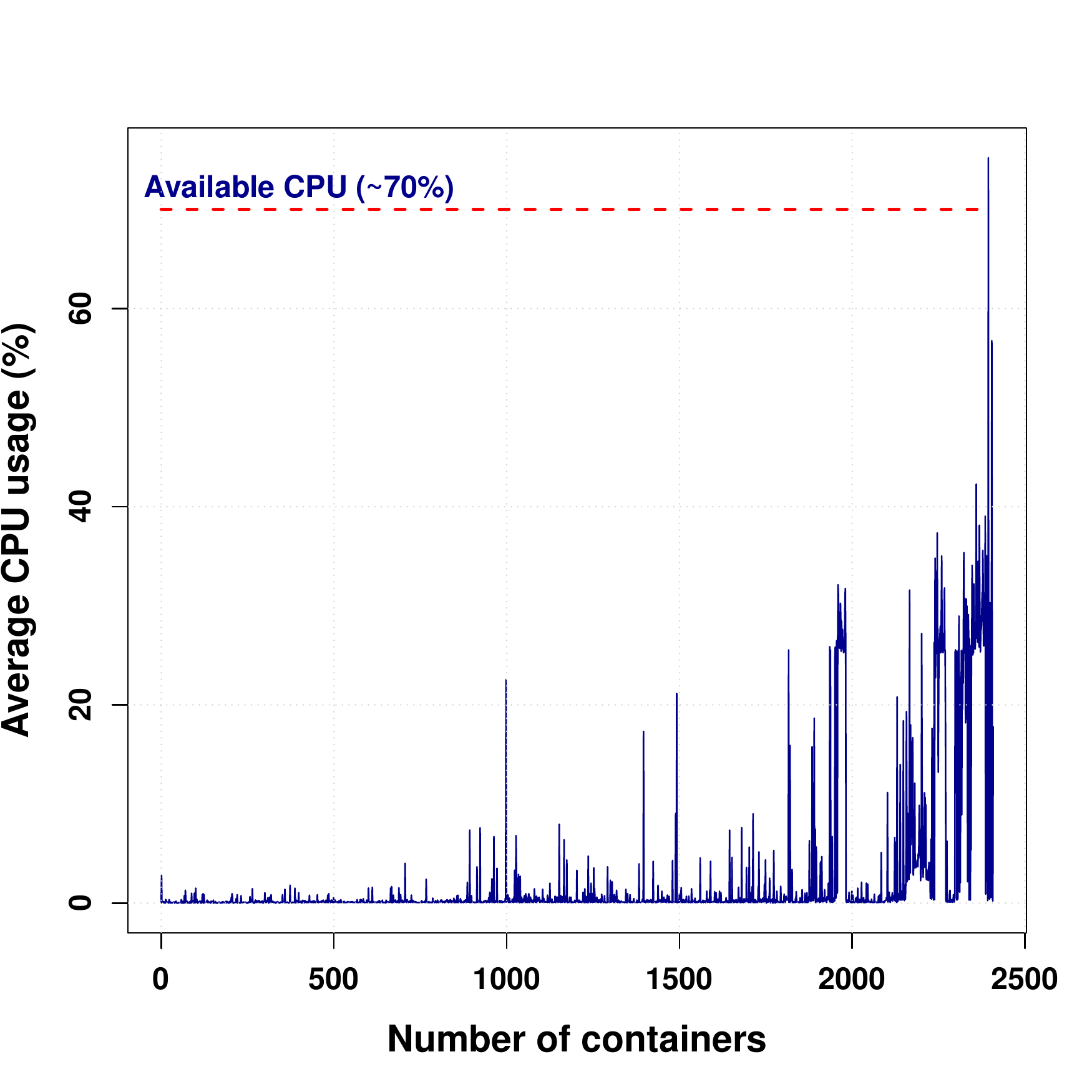} 
	\caption{\scriptsize{CPU usage}} 
    	\label{result:cpu} 
  \end{subfigure} 
  \begin{subfigure}[b]{0.24\linewidth}
    	\centering
    	\includegraphics[width=1\textwidth]{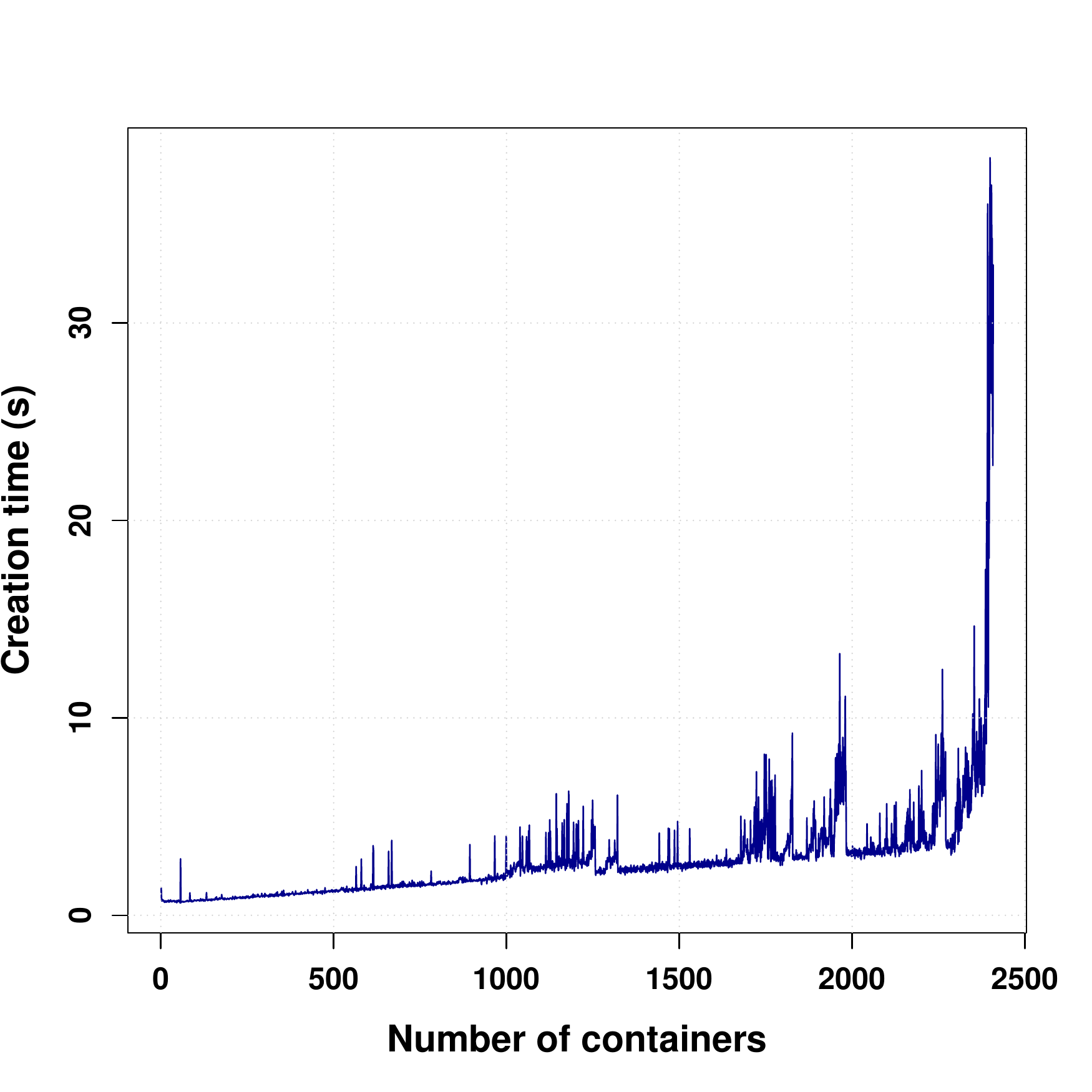} 
	\caption{\scriptsize{Creation time}} 
    	\label{result:time} 
  \end{subfigure} 
   \begin{subfigure}[b]{0.24\linewidth}
    	\centering
    	\includegraphics[width=1\textwidth]{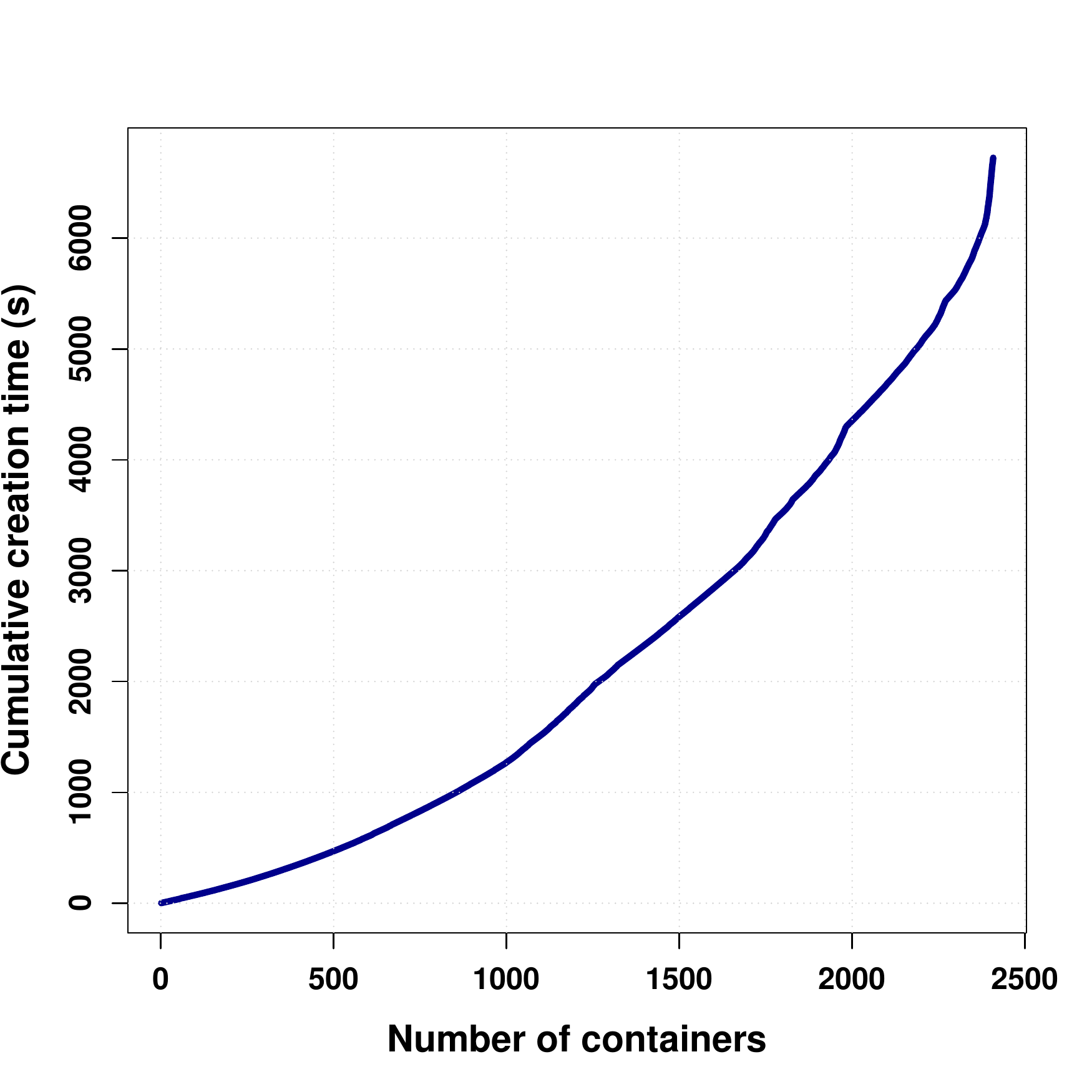} 
	\caption{\scriptsize{Cumulative creation time}} 
    	\label{result:cumulativetime} 
    	 
  \end{subfigure} 
\vspace{-2ex}
 \caption{Number of web servers on a single PI}
\label{result:benchmark}
\end{figure*}

\vspace{-3ex}

 \begin{figure*}[ht] 
  \begin{subfigure}[b]{0.33\linewidth}
    	\centering
    	\includegraphics[width=0.85\textwidth]{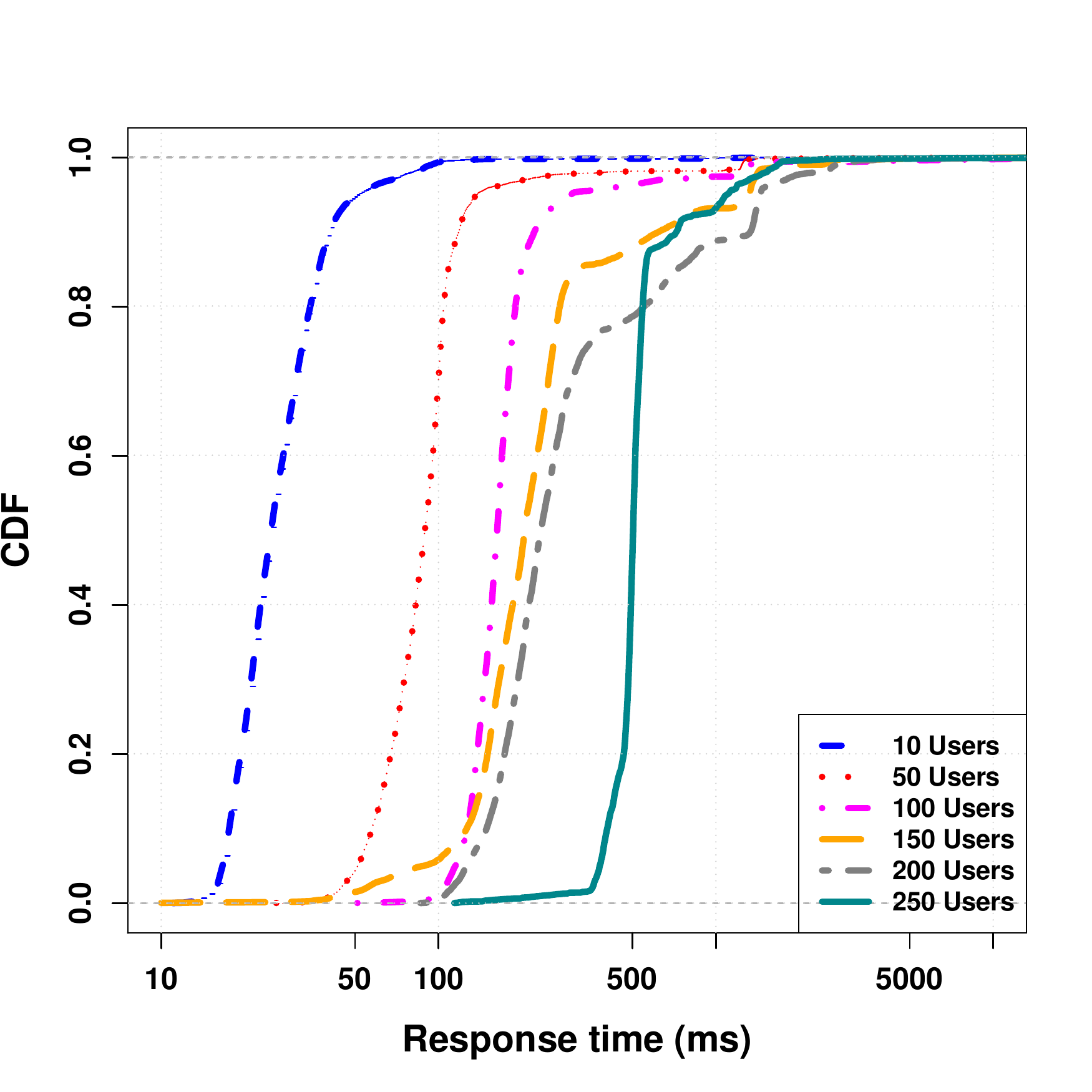}
    	 \caption{\scriptsize{CDF response time }} 
    	\label{result:cdf_delay}
	  \end{subfigure}
  \begin{subfigure}[b]{0.33\linewidth}
    	\centering
    	\includegraphics[width=0.85\textwidth]{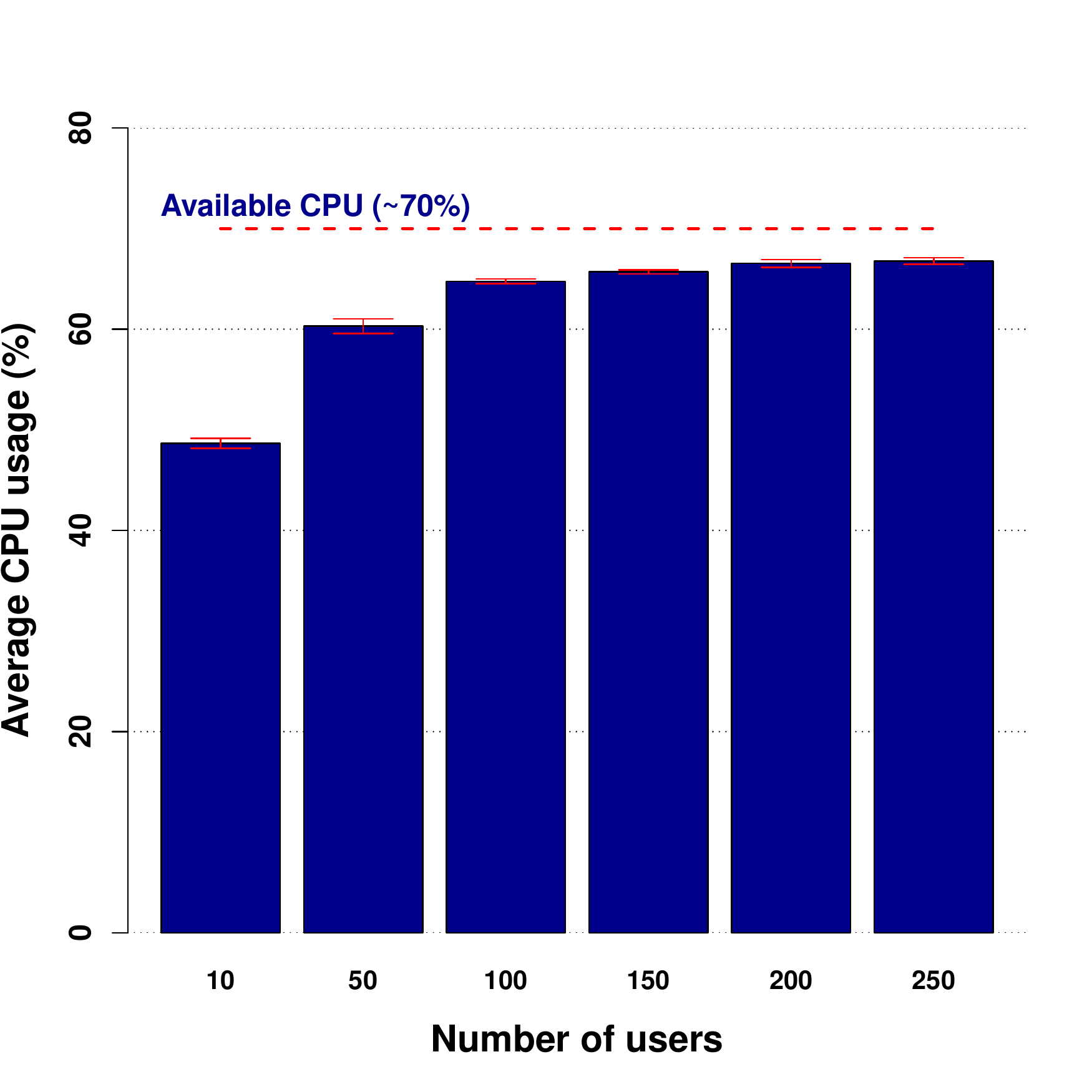} 
	\caption{\scriptsize{CPU utilization}} 
    	\label{result:cpu_utilization} 	
  \end{subfigure} 
  \begin{subfigure}[b]{0.33\linewidth}
    	\centering
    	\includegraphics[width=0.85\textwidth]{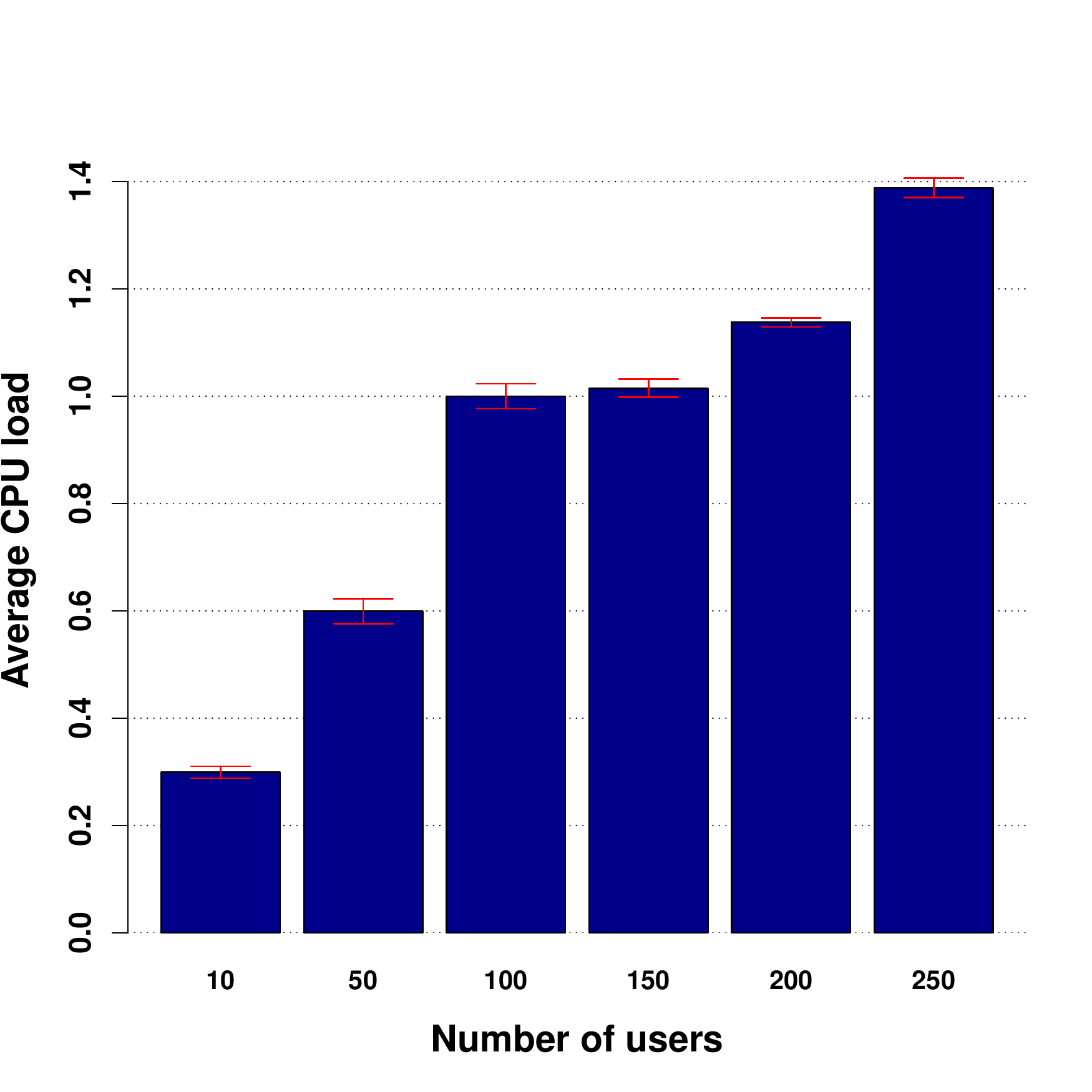} 
	\caption{\scriptsize{CPU load}} 
    	\label{result:cpu_load} 
	 \end{subfigure} 
\vspace{-2ex}
 \caption{Stress test with multiple requests}
\label{result:load_test}
\end{figure*} 
	 
\vspace{2ex}
\subsection{Scaling up the number of deployed containers within a PI}
The first evaluation is to explore the maximum number of containers that could be operated concurrently over a single PI. 
To scale up the containers, the docker image could be prepared as small as possible to minimize memory footprint. Consequently, the memory allocation for kernel to handle a web server process can be optimized. For this, we use a nano web server image\footnote{https://github.com/hypriot/rpi-nano-httpd} developed in assembly code (size is less than 90 KB (included index.html and a small jpg file)). To benchmark the capability of a PI (PI 2), we base our evaluation on memory consumption, CPU utilisation and the creation time (time taken to create a container) by using {\it sysstat}, a collection of performance monitoring tools for Linux . In our first attempt, we were able to spin up only 37 containers, even though the memory usage and CPU utilisation were only 40\% and 18\% respectively. We then hacked the Docker daemon, to scale up the deployed containers to 2408. The procedures that we used is summarised below:

\begin{itemize}
\item{\emph{Tweak docker command:} We disabled some docker options that are unnecessary for running a web server. Especially, running with a dedicated IP stack per container involve a huge resource usage, so we run each container with $net=host$. We also disabled log driver ($log-driver=none$) to let docker use less resources.}
\item{\emph{Unlock docker limit:} Linux distributions use systemd to start the Docker daemon. We customised some parameters to unlock the limitation of maximum number of running containers. The number of processes (LimitNPROC) and number of queued signals (LimitSIGPENDING) were set to infinity.}
\item{\emph{Tweak docker configuration:}With the default configuration, docker daemon is run with many options which is not required for a simple web server such as IPv6, proxies, IP forwarding, log-driver. We disabled all these options to reduce the memory usage of docker daemon.}
\end{itemize}

The results for our optimised docker are depicted in Figure~\ref{result:benchmark} where we were able to scale up the number deployed containers to 2408. Specifically, memory usage is a key factor that limits the capability of running the containers on a PI. As shown in Figure~\ref{result:mem}, the memory usage increases gradually when a container is added. The initial memory usage before creating the first container was about 98 KB (9.89\%). We hit the limit of 2408 containers where there is no space for available memory. 

Figure~\ref{result:cpu} shows the average utilisation of the quad-core CPU of the PI. Over the first thousand of deployed containers, the average CPU usage is very low (about 0.2\%) with a few spikes. However as expected, the CPU usage increases significantly in the high load state (when the number of containers is larger than 2000). As mentioned in Section~\ref{Benchmarking the drone}, the quadcopter consumes about 30\% of CPU resources. The available space can be around 70\% which is sufficient to provide multiple services with Docker containers. Hence when the quadcopter is flying, we will not be able to have 2408 containers running in parallel. However, a \cloudrone can still support a large number of concurrent containers. 

Figure~\ref{result:time} depicts the results of creation time where each point denotes the time it took for the nth container to start up. The creation time for each container varies from 0.62 s to 38.37 s depending on the current CPU load and memory usage. In addition, we also plot the cumulative creation time of the 2408 containers (Figure~\ref{result:cumulativetime}). The PI spent approximately 1 Hr and 50 minutes to spin out 2408 containers. On average each container requires about 2.79 s to start up the web server. 

{\it Key takeaway message}: A single PI (PI 2 model B) can support significant amount of concurrent lightweight services.
\vspace{-2ex}

\subsection{Scaling up the number of users accessing a single service}
In order to investigate the feasibility of using lightweight containers as a platform for the \cloudrone, we aim to evaluate the scalability of each of these containers running on a single PI while serving a large number of requests. We deploy a minimal static web server using httpd docker base image using a similar configuration as the experiments in the previous section. 
The benchmarking scenario represents the {\it Cloudrone's} operation where users on the ground can access the services provided by the {\it Cloudrone} through a wireless interface. Using the {\it Ab} - Apache HTTP server benchmarking tool\footnote{https://httpd.apache.org/docs/current/programs/ab.html}, we conduct stress tests on the {\it Cloudrone} while scaling the number of concurrent users from 10 to 250. The total number of request was set as 10000 transactions per experiment. For instance, in case of 10 users, 1000 of requests were sent by each user. The measured RTT via ping tests with a clear line of sight at 50 feet was between 8ms-10ms. 

Figure~\ref{result:cdf_delay} illustrates the CDF of response time from the web container running on PI while varying the number of simultaneous users accessing the service. As shown in the figure, a single deployed docker container is capable of serving a large number of  concurrent users. 
As expected, the average response time increases when the number of concurrent users is scaled up. Figure~\ref{result:cpu_utilization} and ~\ref{result:cpu_load} plot the CPU utilisation and CPU load of the PI using \emph{sysstat} tools. The CPU utilisation increases almost 20\% when the number of users increase from 10 to 250. This increases the response time for the processes. Even though the utilization has not reached the capacity of CPU, processes still run slower as the CPU load increases. Figure~\ref{result:cpu_load} shows the average CPU load of the PI (sampled in one min intervals).  As the arrival rate of user requests increases, the amount of computational work need to process also increases. This has impact on the response of time(Figure~\ref{result:cdf_delay}).  



{\it Key takeaway message}: A Docker container running on a single PI (PI 2 model B) can support significant amount of concurrent users.

%
%
\vspace{-1ex}
\section{Discussion}
\subsection{Scalability Challenges}
Our preliminary benchmarking demonstrates that the PI is capable of functioning as a great micro cloud platform. Our benchmarking was carried out using a lightweight web server serving a lightweight webpage.  It is important that the scalability of the \cloudrone should also be tested with heavier web servers serving applications such as Openstreetmaps. Each application will have different memory and CPU requirements and hence the number of containers that can be instantiated will vary depending on the type of applications which directly influences the size of the container (e.g., packages, data, library).

Another scalability challenge is to support a larger number of users within an area.  Increasing the number of users causes an adverse influence on the response time which will cause service degradation. We envision, the different services will be provided by a swarm of drones and hence appropriate load balancing using techniques such as application layer anycast~\cite{Zegura:2000:AAS:348744.348748} could be used.

\subsection{Service Retrieval}
{\it Cloudrones} have two main challenges in terms of providing a reliable service. First, mobility poses a critical challenge. During mobility, ongoing sessions may break and sessions need to be reestablished. Second, considering the distributed nature of the services across a mesh of drones, identifying the location of the service across a mobile adhoc network is challenging. To solve the latter, techniques such as Multicast DNS (mDNS)~\cite{mdns} or application layer anycast could be used~\cite{Zegura:2000:AAS:348744.348748}. Another potential way to mitigate the problem of mobility and service discovery is to explore new architectures that utilise Information Centric Networking)~\cite{Jacobson:2009}. ICN architectures such as Named Data Networking (NDN)~\cite{Jacobson:2009} or SCANDEX~\cite{Sathiaseelan:2015:SSC:2753488.2753490} decouple the content from the location thus removing the need for the current end-to-end client server model such that the service and/or content can be served directly by any host that currently has the service/content. ICN thus integrates the provisioning of content with the locationless notion of information delivery in ICN allowing different flavours of caching, from on-path caching to edge caching through a farm of surrogate micro servers running on the {\it Cloudrones} that can be quickly integrated into the overall (ICN) routing fabric without the need for DNS redirection or other solutions of the current Internet. This inherently addresses the issues of mobility and reliability. As ongoing work, we are in the process of developing an ICN architecture for {\it Cloudrones} via the EU UMOBILE project~\footnote{http://www.umobile-project.eu/}.

\subsection{Deployment Issues}
Although {\it Cloudrones} demonstrate excellent potential for deploying localised service infrastructures in areas where access to services are crucial but are beyond reach - there are still major challenges that need to be surmounted. 

Drones such as quadcopters have reduced flight times due to battery life. We envision this situation will change in the near future with better innovations in battery design and production or innovations in alternate sources of power e.g. hydrogen powered drones have flight times upto two hours~\footnote{http://www.bbc.co.uk/news/technology-35890486}. {\it Cloudrones} also need not be in the air for their entire flight duration to provide access to its services. We envision that {\it Cloudrones} can be flown to an area and then can provide it's localised services from the ground (ideally powered by an energy source on the ground).

There are tight regulations in flying drones such as quadcopters. These rules have been laid out by Civil Aviation Authority (in the case of UK)\footnote{https://www.caa.co.uk/drones/}. Hence these rules should be adhered to and in some cases may be restrictive e.g. land or fly in a congested area. However, we believe, \cloudrone deployments will fall under the commercial aerial work, and hence special permission from the aviation authority will be required to fly.

\vspace{-1ex}

\section{Conclusions}

In this paper, we present \cloudrone - a preliminary idea of deploying a lightweight micro cloud infrastructure in the sky using indigenously built low cost drones, single board computers and lightweight Operating System virtualization technologies. We describe an initial design of the \cloudrone and provide a preliminary evaluation of the proposed system mainly focussed on the scalability issues of supporting multiple services and users. As part of future work, we plan to conduct large scale evaluation trials benchmarking the \cloudrone performance while in the air (in terms of throughput, latencies and energy) across a wide set of scenarios. We are also in the process of integrating Docker with NDN and performance benchmarks will be carried out. Finally, the current \cloudrone design does not fly autonomously and hence is strictly limited in terms of distance it can cover without manual intervention. As part of future work, we intend to build autonomous flying capabilities.


\end{document}